\documentclass{ws-jai}
\usepackage[flushleft]{threeparttable}

\usepackage{amssymb}
\usepackage{amsmath}
\usepackage{gensymb}

\begin{document}


\markboth{Deshpande and Lewis}{Mitigation of RFI from IRIDIUM Satellite Signals}

\title{IRIDIUM SATELLITE SIGNALS: A CASE STUDY IN INTERFERENCE
CHARACTERIZATION AND MITIGATION FOR
RADIO ASTRONOMY OBSERVATIONS}

\author{Avinash A. Deshpande$^{1,2}$ and  B. M. Lewis$^{2}$}

\address{
$^{1}$Raman Research Institute, Sadashivanagar P.O., Bangalore 560 080 INDIA\\
$^{2}$NAIC/Arecibo Observatory, HC3 Box 53995, Arecibo, Puerto Rico 00612
}

\maketitle

\corres{$^{1}$Avinash Deshpande desh@rri.res.in}


\begin{abstract}
Several post-detection approaches to the mitigation of radio-frequency
interference (RFI) are 
compared by applying them to the strong RFI from
the Iridium satellites. 
These provide estimates for the desired signal in the presence of RFI,
     by exploiting distinguishing characteristics of the RFI,
 such as its polarization, statistics, and periodicity.
Our data are dynamic spectra with full
Stokes parameters and 1 ms time resolution.
Moreover, since most man-made RFI is strongly polarized,
we use the data to compare its unpolarized component with its Stokes I.
This approach on its own reduces the RFI intensity by many tens of dBs.
A comprehensive approach that also recognizes
non-Gaussian statistics, and 
 the time and frequency structure
inherent in the RFI permits exceedingly effective post-detection excision
provided full Stokes intensity data are available.

\end{abstract}

\keywords{methods: data analysis, radio lines: general, radio-frequency interference, 
techniques: polarimetric}

\section{INTRODUCTION}
\noindent 
Some sources of radio frequency interference (RFI) are inescapable. 
While radio astronomers can minimize the
effects of many terrestrial sources by placing their telescopes 
at remote sites, none can escape from RFI generated by satellite 
transmitters, such as those of the Iridium System. So astronomers 
are now studying a variety of diverse approaches to mitigating 
the effects of RFI on their observations (Ellingson 2005, and references therein).
The {\it pre-detection} approaches to RFI excision relate 
to those applied to time sequence of signal {\it voltage}; 
that is {\it before} the stage of {\it square-law detection} 
or translation to intensity, such that signal phase information
is still available. 
On the other hand, the post-detection RFI mitigation
methods are relevant to a more commonly available form of data sets
of {\it intensity}, estimated in general
as a function of time and frequency, that is dynamic spectra with 
desired temporal and spectral resolutions and spans.
More broadly, the {\it post-detection} data refer to 
any quantity proportional to average intensity,
such as estimates of cross-correlation of signal voltage sequences from two elements
of an interferometer or from two orthogonal polarization feed antennas, which 
preserves only the {\it relative} phase between the correlated signals.

Some of these exploit 
particular characteristics of specific sources of RFI, such as its 
location, if from a geostationary satellite, or its polarization.
Indeed, since man-made signals are highly polarized, whereas 
the inherent nature of most astronomical signals is unpolarized, 
Deshpande (2005) proposed the use of a mitigation technique based on 
the unpolarized, relatively RFI-free signal.  In this paper, we 
compare and contrast this approach with several other post-detection
approaches to the identification and mitigation of RFI originating 
from the Iridium System.

\section{OBSERVATIONS}

The International Telegraphic Union (ITU) granted 
the Radio Astronomy Service (RAS) primary status 
in the 1610.6-1613.8 MHz band in 1992 to observe 
the 1612.235 MHz spectral line emission from the 
hydroxyl molecule (OH). This is typically emitted 
by OH/IR stars\footnote{
These are Asymptotic Giant Branch (AGB) stars that 
show strong OH maser emission and are unusually bright 
in near-infrared (IR).} (see Lewis, Eder \& Terzian 1985)
 as a pair of narrow features, with 
the allocated band sized to allow for Doppler shifts 
of the emission, as well as guard-band separation 
from ITU Services using adjacent spectrum. 
One such is the Iridium L-Band system, which presently 
uses a 1618.85-1626.5 MHz allocation. But this system 
also produces a comb of RFI, with a characteristic 333 kHz 
spacing (often with 8 times finer sub-spacing), 
extending well beyond its licensed band. 
The $\sim$1 Jy intensity of this comb in the RAS band 
for most Observatories is comparable with the signal 
from many of the brighter OH/IR stars. No pre-launch 
simulation available to the ITU or to radio-astronomers 
gave any hint of the existence of this noxious artifact 
when the System was granted spectrum. 
Hence the need now for RFI mitigation.

This study was made using the highly sensitive 
305 m Arecibo telescope, which has 80 dB of forward gain 
and thus a narrow main beam. Accordingly the 66 active, 
low-earth-orbit satellites from the Iridium System are 
only generally seen at Arecibo in distant sidelobes: 
that lessens their impact on our observations. 
As most of Iridium's customers near Puerto Rico are on yachts, 
local terrain generally screens us from their up-link signals. 
Moreover as the Iridium System is lightly loaded in the 
Caribbean it emits few inter-modulation products in 
the signal received at Arecibo from interactions within 
a satellite's own components.

Iridium uses a frequency multiplexed -- time multiplexed 
operational mode. This gives two helpful features, 
as the signal is strongly polarized 
(more specifically, has right-hand circular polarization), 
and, more unusually, 
has a satellite down-link time-multiplexed in exactly 
the same band as the phone handset up-link signal. 
Each Iridium satellite operates on a 90 ms cycle, 
with half assigned to the up-link, and half to the downlink. 
That allows the folding of our data at a secondary period 
of 180 ms, or twice the basic cycle. The timing operations 
of the entire Iridium constellation is governed by the 
System's most intense signal, the 1626 MHz clock synchronization signal.

\begin{figure}[h]
\begin{center}
\includegraphics[width=6.0in,angle=-90]{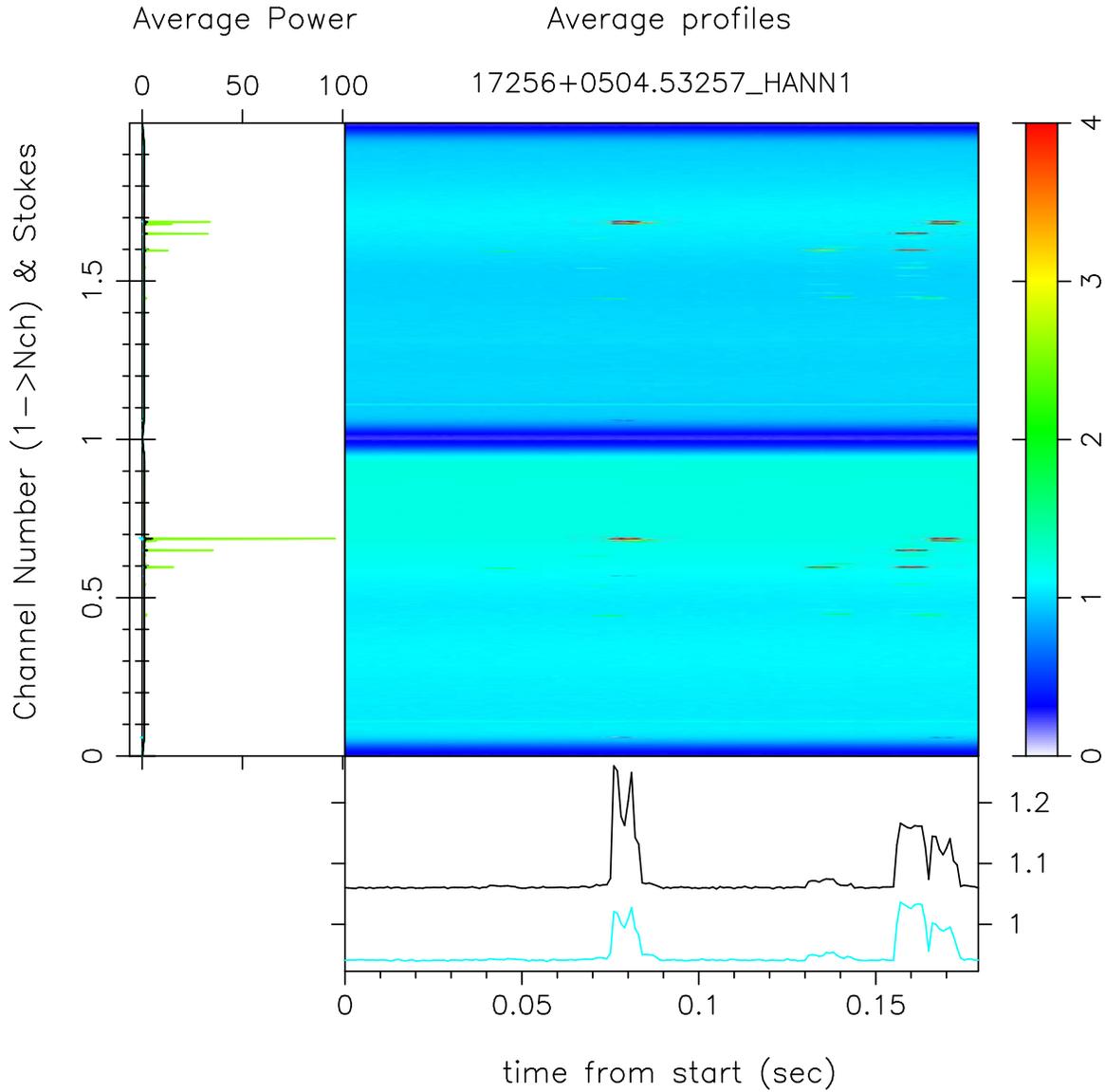} 
\end{center}
\caption{Example of typical 180 ms long dynamic spectra for 
the dual linear polarization channels (X and Y, in the upper and 
lower halves of the main panel), each observed across a 25 MHz wide band
centered at 1622 MHz.
The vertical axis labels for both the left-side panel and the central panel
indicate a normalized channel number (normalized by number of channels across 25 MHz)
so as to correspond to the number of polarization (or Stokes) products displayed
(two in the present case). 
The two line plots in the bottom panel show the corresponding time profiles
of the band-averaged intensities in X and Y, respectively. 
The intensities
(indicated along the vertical axis of the bottom panel, for the
color plot, as well as along the horizontal axis of the left panel) 
are consistent across the panels, but are in an arbitrary scale.
The left side
panel has the spectral profiles for the maximum (green), 
the minimum (sea-blue), and the average intensities. 
Here, the latter two are
almost invisible in the scale dictated by the maximum intensity apparent
due to Iridium RFI. See the main text for further details.}
\label{aba:fig1}
\end{figure}

\begin{figure}[h]
\begin{center}
\includegraphics[width=6.0in,angle=-90]{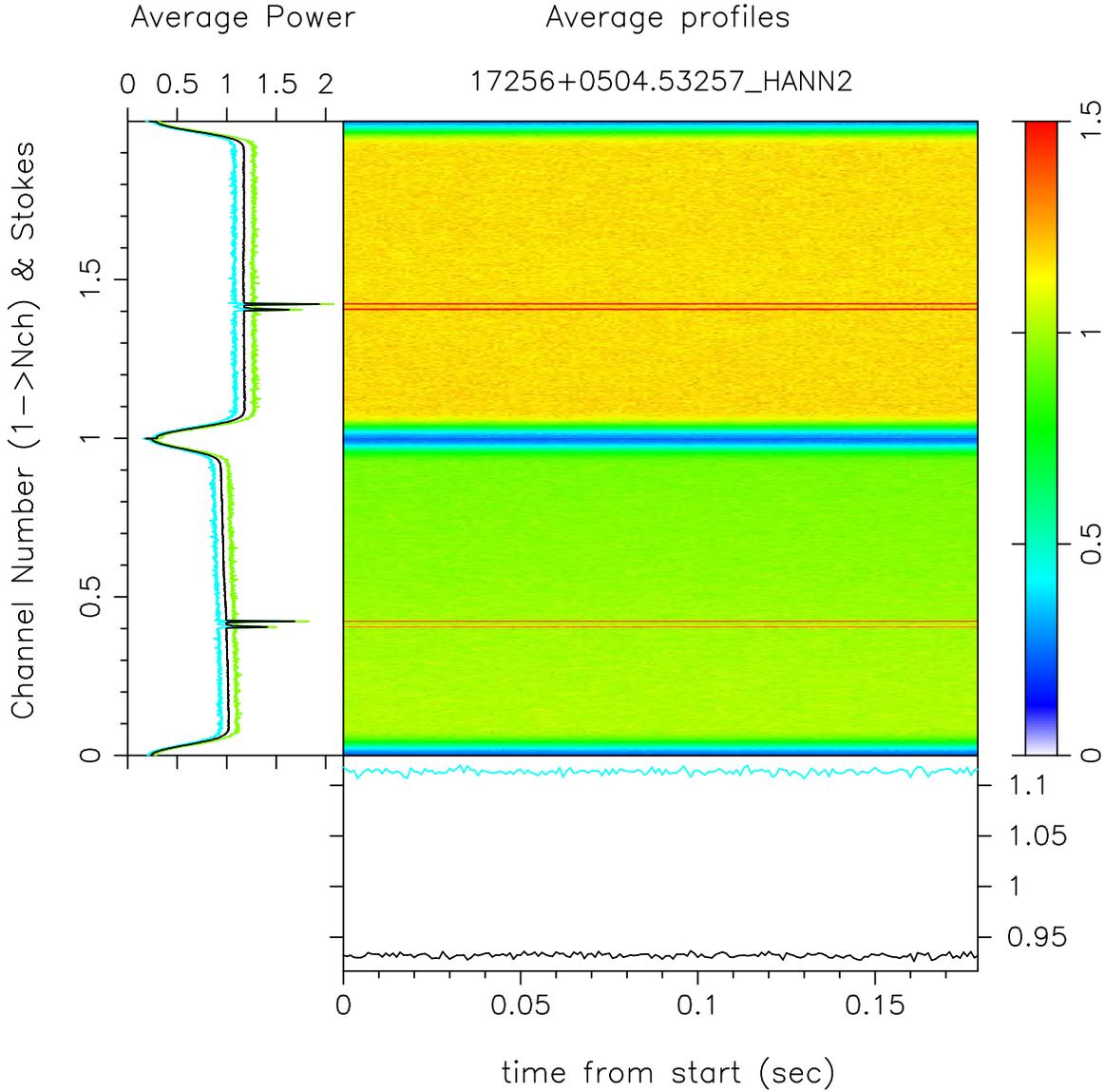} 
\end{center}
\caption{Similar to Fig. 1, but for a bandwidth of 3.125 MHz
in the RAS band (centered at 1612.5 MHz), which is narrow
and so has a correspondingly better spectral resolution, providing
a more resolved view of the line features from the star.}
\label{aba:fig2}
\end{figure}

\begin{figure}[h]
\begin{center}
\includegraphics[width=6.0in,angle=-90]{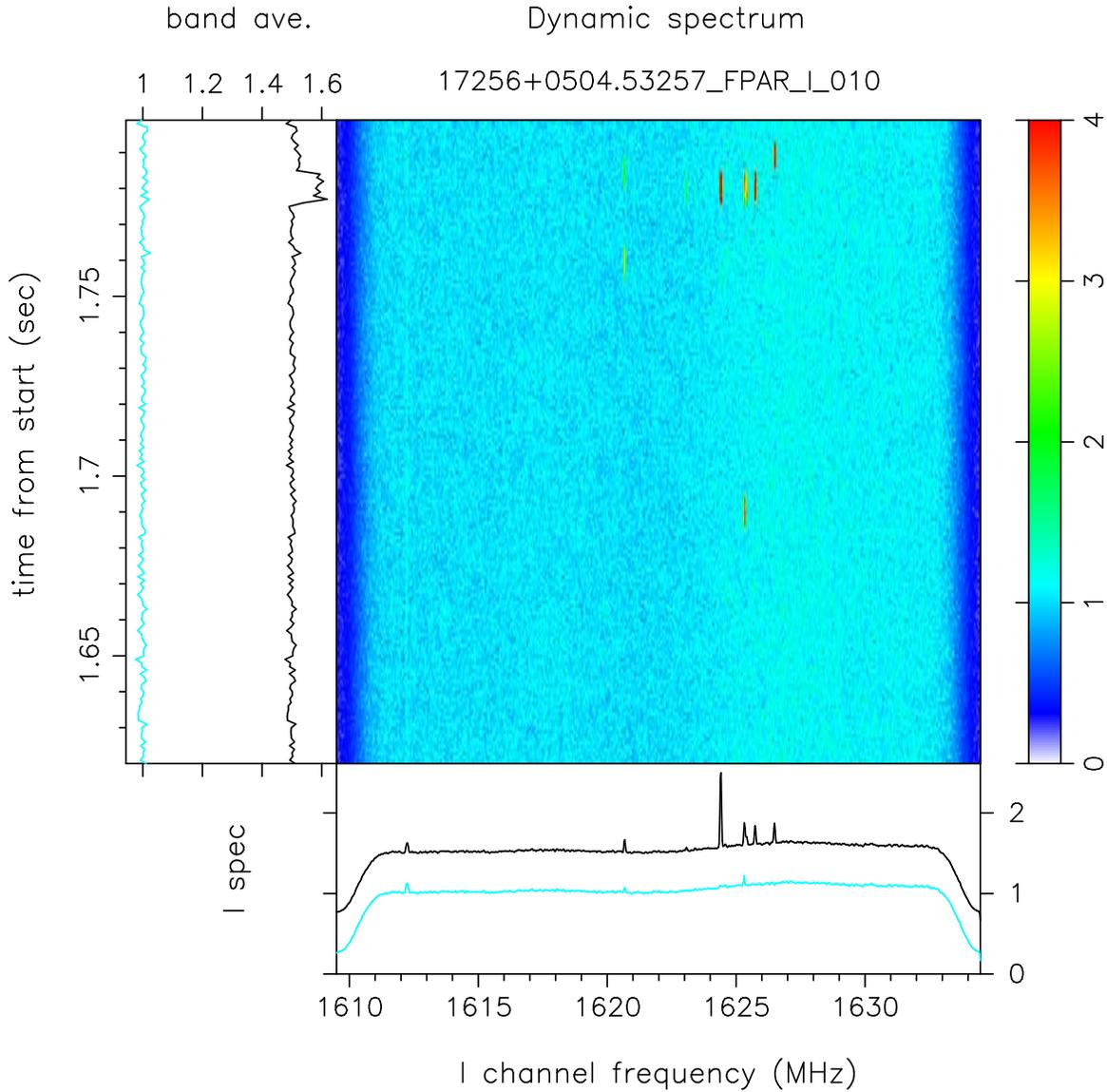} 
\end{center}
\caption{A 180 ms sequence of Stokes I spectra taken at 1 ms 
intervals, together with their arithmetic (black) and robust (coloured) 
means on each axis. Only the $\sim$1612 MHz feature comes from the star.
The black line plots are deliberately offset by 0.5 for clarity.}
\label{aba:fig3}
\end{figure}

\begin{figure}[h]
\begin{center}
\includegraphics[width=6.0in,angle=-90]{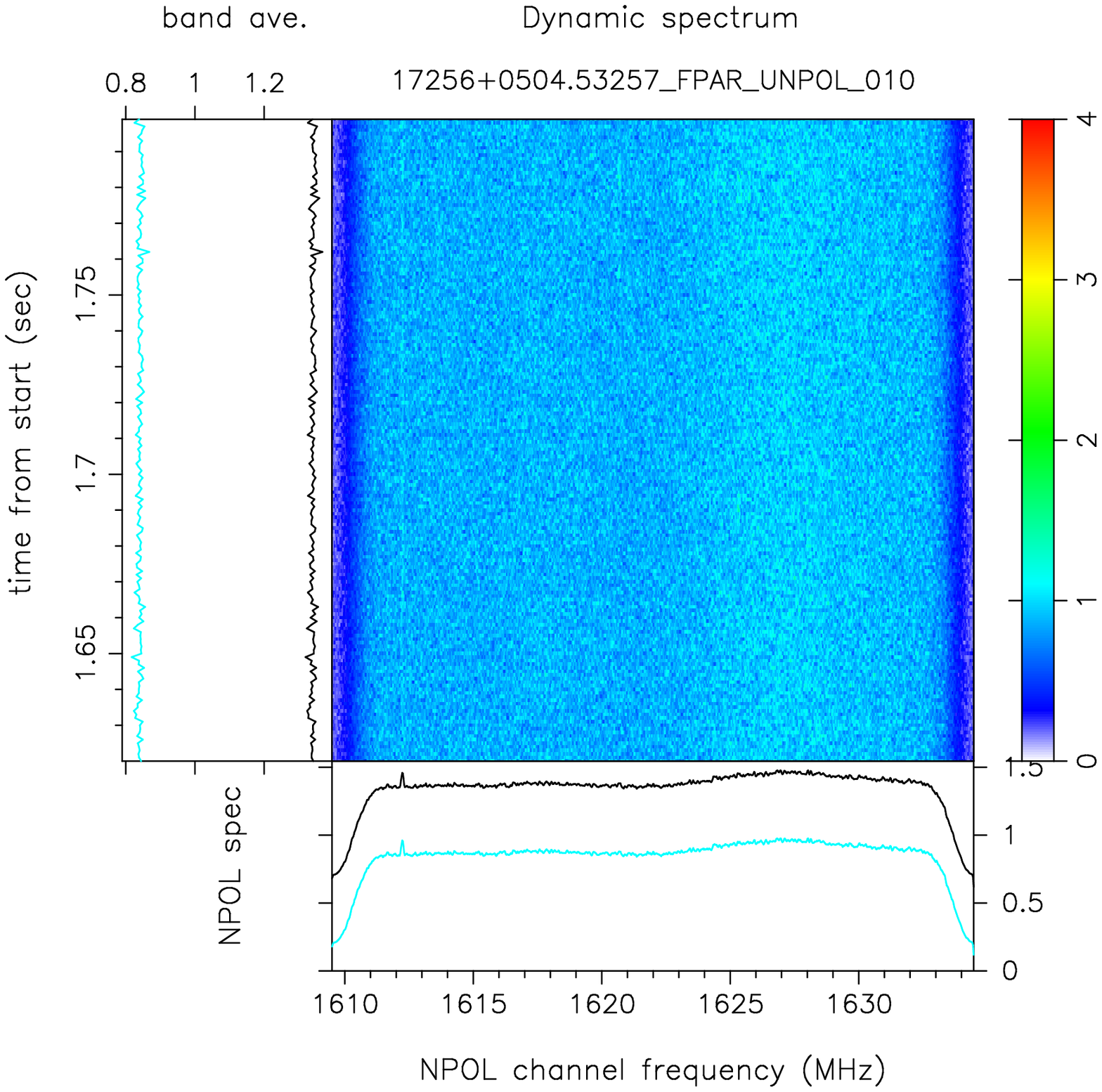} 
\end{center}
\caption{Dynamic spectrum similar to that in Fig. 3, 
but now for the unpolarized component $I_u$,
i.e. after removing the polarized contribution from the Stokes I,
and thus mostly free of RFI.}
\label{aba:fig4}
\end{figure}

Our aim here is to explore and quantify how well the
effect of RFI  can be attenuated while estimating our desired signal.
For this purpose, we need to clearly recognize aspects in which 
our signal differs from contaminating RFI, based on which
the two can be separated. How well such separation is achieved 
in practice in the presence of inherent uncertainty in estimation
of relevant quantities, will decide the effectiveness of 
the RFI mitigation.
Here, we exploit distinguishing characteristics in aspects
such as polarization, statistics, and periodicity, in addition to
differences in structure across time and frequency.
In the discussion to follow, we compare the effectiveness of 
these criteria when applied in isolation, as well as in combination.

Our data were acquired with full Stokes (I, Q, U, V) 
parameters as high time-resolution, single-dish spectra.
Dynamic 1024 channel spectra were recorded every millisecond 
using the 9-level sampling mode of an auto-correlator 
simultaneously in both the Iridium band, using a 25 MHz bandwidth 
centered at $\sim$1622 MHz, and in the RAS band using a 3.125 MHz 
bandwidth with proportionately finer spectral resolution. 
Figures 1 and 2 show examples of typical dual-polarization
channel dynamic spectra for the broad and the narrow bands, respectively, 
obtained from the recorded auto-correlations after Hanning smoothing. 
As can be seen from Fig. 1,
the peak intensities of Iridium RFI in some of the spectral
channels (see the green peaks in the left side panel)
can be orders of magnitude larger than the average spectral
contribution dictated by the system temperature. Fortunately,
the spectral dilution (typically by the number of 
spectral channels, which is here 1024) leads to 
a total power (or band-averaged power) increase of only 10-20\%. 
The RAS band
spectra in Fig. 2 (shown for the same time interval as in Fig. 1),
on the other hand, are contrastingly mostly
free of the Iridium RFI. 
This does indeed document the absence any  RFI into the RAS 
band in the form of a 333 kHz comb, thanks
to the Iridium system being lightly loaded in the Caribbean region.
Hence, we will focus hereafter on the mitigation of RFI in the Iridium band itself,
although the same techniques are equally applicable to RFI mitigation
in the RAS band.

The data were checked for inter-modulation products and 
gain-compression effects induced in the receiving system 
by the Iridium signal by generating correlation maps (following
Deshpande 2005). 
These accumulate the net temporal correlation between fluctuations 
in every possible combination of spectral channels from every 
possible pairing of spectra: they are produced as 
cross-correlations between spectra in the native linears, 
in I, in the unpolarized flux, $I_u$ ( = $I - \sqrt{Q^2 + U^2 + V^2}$), 
as well as between the two
observed bandwidths, and as autocorrelations of spectra. 
After folding 2 minutes of data at the period of the
Iridium clock cycle, the only Iridium artifacts in our 
RAS band data are momentary gain compression episodes 
(see Deshpande \& Lewis 2005). 
Hence, it is not surprising that in this band, 
the statistical measures (such as the average spectrum and 
the rms values of the noise) during the cycle phases 
corresponding to the Iridium RFI are found to be
very similar to those for the rest of the cycle (without the RFI).

\begin{figure}[h]
\begin{center}
\includegraphics[width=6.0in,angle=-90]{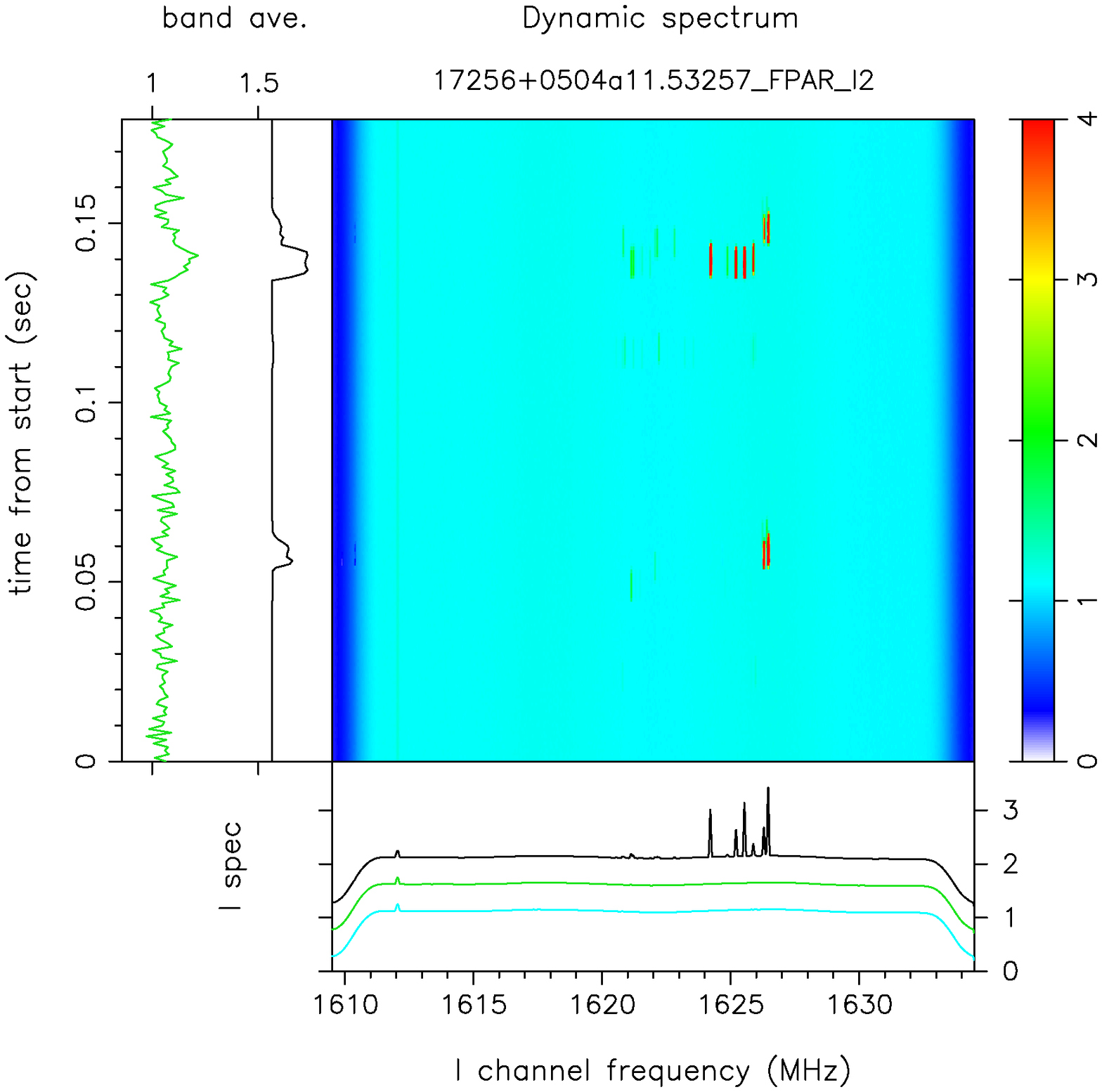} 
\end{center}
\caption {Stokes I, gain-calibrated data (black) folded 
for two minutes at the 180 ms (i.e. two-cycle) Iridium period. 
These are compared to the robust (green) and the ``RFI-free" $I$ 
average from the first 40 ms of each Iridium cycle (sea-green), 
after limiting its constituents to band means that are within 
3$\sigma$ of the overall mean. The band averages for each
millisecond of the Iridium cycle are shown on the left, 
with $I$ offset by 0.5 and the
robust mean, $z$, displayed as $(z(t) - \left<z\right>)*100 + \left<z\right>$
(so as to amplify the deviations for ready visibility).}
\label{aba:fig5}
\end{figure}

The intensity of the Iridium clock synchronization signal (at 1626 MHz)
generates some ringing in the adjacent spectrum,
despite our nine-level sampling, so our data is always 
Hanning smoothed, as, for instance, in Figures 1,2,3. 
Detailed channel by channel examination of the data, 
shows that there is still a noticeable under-correction of data in
frequency bins immediately before and after the synchronization 
impulse, and we find that this exaggerates the
residuals in comparisons with $I_u$: data in Figures 5 \& 6 
are therefore Hanning smoothed twice. The bandpass gain
calibration is applied to the linear polarization data before 
computing Q \& $I_u$ from the Stokes I, Q, U, \& V of
each channel of each spectrum as $I_u$ = $I - I_p$ (where 
the polarized component is estimated as $I_p$ = $\sqrt{Q^2 + U^2 + V^2}$),  
which reduces the residuals from $I_u$.

\section{MITIGATING}

\begin{figure}[h]
\begin{center}
\includegraphics{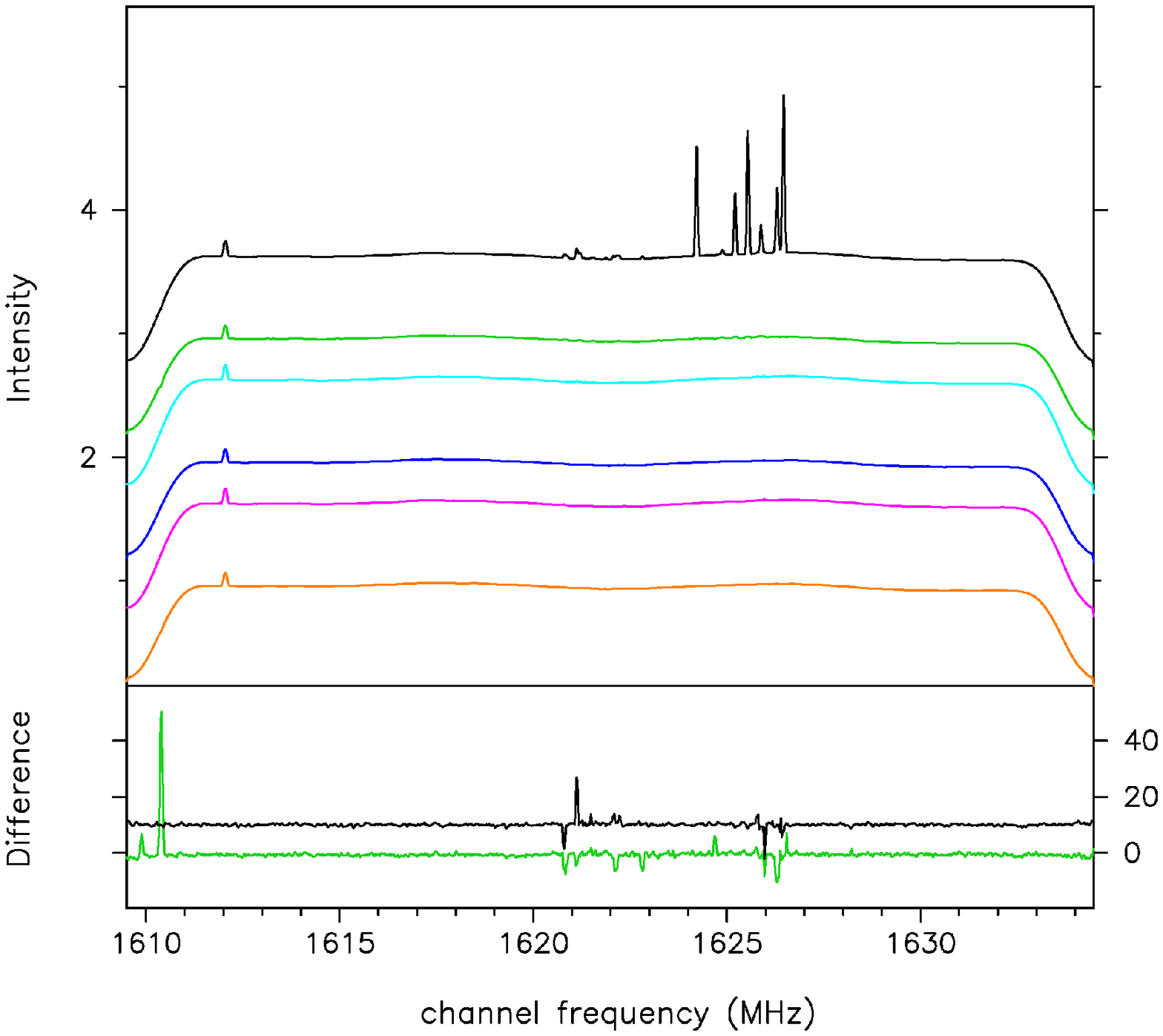}
\end{center}
\caption{The two-minute folded mean intensity spectra in descending order:
(1) arithmetic average Stokes $I$ (black); 
(2) $I_u$ (green); (3) robust mean of $I$ (sea-green); 
(4) robust mean of $I_u$ (blue); (5) ``RFI-free" mean (violet); 
(6) $I_u$ mean from the RFI-free set (brown). 
The intensity scale in the top panel 
is in normalized units with respect to the system temperature
$T_{sys}$. Successive spectra are deliberately offset along 
vertical axis in multiples of 0.5 units in normalized intensity
for providing clear view of their individual spectral variations. 
The second (bottom) panel shows the differences (3)-(5) in black \& (4)-(6) in green
on a magnified scale, in units of expected rms 
(i.e. standard deviation of noise).}
\label{aba:fig6}
\end{figure}

All of the Figures, except Fig.2, use 25 MHz bandwidth data obtained 
towards the OH/IR star IRAS 17256+0504 on 9th September 2004 
(azimuth and zenith angle of $\sim$330$\degree$ and $\sim$15$\degree$, 
respectively).
Fig. 3 shows a typical, 180 ms (two-cycle) sequence of 
Stokes I spectra with the 
OH/IR star at $\sim$1612 MHz and Iridium's signal between 1620-1627 MHz. 
The mean side-lobe response of the strongest Iridium Stokes I
feature here, which is only seen for $\sim$5\% of the time, 
has $\sim$7 times the intensity of our strong OH/IR star which is seen
with the main beam. So Iridium can easily saturate an astronomical 
receiver (for examples, see Deshpande \& Lewis 2005). 
Perhaps the simplest mitigation approach is to use 
the robust mean\footnote{ Robust mean and robust rms refer to 
respectively the arithmetic mean $<x>$ and the standard deviation $\sigma$ 
associated with a distribution of an ensemble of values
after excluding identifiable outliers from the ensemble. 
Outliers are the samples with values outside the expected range
$<x> \pm \gamma \sigma$, where $\gamma$ defines the threshold
in units of $\sigma$ on either side of the mean. For an ensemble of size 
$N$, the $\gamma$ can be specified such that the probability of
value being outside the above range is less than 1/N, for a
Gaussian distribution. An iterative procedure refining the ensemble
is expected to converge in a few iterations.},
which only discriminates against obviously 
non-Gaussian components in a series. It is computed from 
(and may modify) the temporal sequence of each frequency channel. 
While the robust mean excises most of the strong RFI in Fig. 3, 
though leaving a clear residue in the average spectrum
circa 1621 \& 1625.5 MHz and a broad, slight one circa 1624.5 MHz, 
the vertical panel shows that it otherwise tracks I. 
The residues from the robust mean in the average spectrum in 
effect arise from integrating up features below the robust 
threshold. On the other hand a similar plot, as in Fig. 4, for the second 
mitigation approach using $I_u$ shows no evident sign of 
an Iridium signal. So $I_u$ over short, one-on-one, 
time comparisons appears to provide better mitigation, 
and its output is improved further by applying a robust mean. 
Yet the efficacy of $I_u$ relies on the estimation of the 
polarized component $I_p$, 
which is both over-estimated (due to direct contribution from
variance of random noise in the $Q$, $U$ and $V$ on squaring)
and least accurate for weak features: indeed the strongest 
RFI features are the most accurately excised, 
as the estimation of the polarized flux is then done with 
the best S/N, while that of the weakest is subject to 
a statistical bias, for which no allowance is made here. 
And $I_u$ also leaves residuals when the {\it apparent}
polarization of RFI is reduced  
as a result of depolarization due to averaging
of different polarization states (such as while folding), 
as Fig. 5 shows at $\sim$1622 MHz, and Fig. 6 for 
features circa 1624 \& 1626 MHz. 
Apparent reduction in fractional polarization 
could result from poor calibration of the Stokes parameters, 
due to uncertainties in the relative (complex) gains, as well as
due to saturation-induced gain calibration errors, if any. 

But folding spectra, 
as Fig. 5 does, permits both weak RFI features below the 
robust mean's threshold and, as the second panel in Fig. 6 
shows, any coherent distortions in the estimation of $I_u$, 
to emerge from the noise. Finally the folded spectra of Fig. 5 
offer a third approach to mitigation, as portions of the 
Iridium cycle without RFI are readily identified from its 
left-hand panel. We select ``RFI-free" spectra by additionally 
requiring that their integrated total power lies within 3$\sigma$ 
of the mean of the set. Figures 5 \& 6 include RFI-free means.

Fig. 6 enables the integrated-up, coherent deviations to be 
exhibited. It shows that much the largest deviation occurs 
in $I_u$ on the ascending edge of the bandpass at $\sim$1610.5 MHz, 
near a resonance in the ortho-mode transducer where the 
polarization properties of the receiver change rapidly. 
Since its specific properties were not addressed here, 
we discount it. All of the other excursions in the difference 
sets occur in the vicinity of evident RFI, with the strongest 
(circa 1621 MHz) linked to weak RFI in both the robust and 
unpolarized means, and the largest linked to the robust means. 
Nevertheless there are no systematic deviations in either approach 
over most of the spectrum. Table 1 shows the attenuation in dB 
achieved with the features in Fig. 6. It should be noted that
the Iridium data stream is more than usually amenable to treatment 
by the robust approach on its own, as its occupancy of its band 
is both episodic and less than 50\%. In cases where strong RFI 
has a much larger fractional occupancy, so there is little 
uncontaminated data, the robust mean on its own will fail, 
whereas that situation is usually greatly improved if $I_u$ 
is appropriate and available.

\begin{wstable}[h]
\caption{Attenuation achieved on the 9 Stokes I features in Fig. 6 
in order of increasing frequency.}
\begin{tabular}{@{}cccccccccc@{}} \toprule
method (dB)           & 1 & 2 & 3 & 4 & 5 & 6 & 7 & 8 & 9 \\ \colrule
robust                & -32 & -33 & -29 & -29 & -34 & -15 & -43 & -19 & -09 \\ 
unpolarized ($I_u$)   & -32 & -22 & -14 & -21 & -18 & -08 & -20 & -08 & -13 \\ 
$I_u$ + robust        & -32 & -28 & -38 & -27 & -28 & -15 & -33 & -18 & -23 \\ \botrule
\end{tabular}
\label{aba:tbl1}
\end{wstable}

Our analysis has been carried through on a data stream 
corrected for the polarized flux estimate on the
time-scale of the shortest integrations, which is usually the worst case. 
This processing could be adjusted to improve on Hanning smoothing, 
and by using statistical bias corrections. It could also be 
better adapted to the needs of spectral-line observers, 
by iterating the analysis on data processed for longer 
integration times generated after synchronously averaging 
the dynamic spectra from each sub-interval within 
the Iridium cycle. On the other hand when only a band-averaged 
power level measurement is wanted, any source of narrow-band RFI 
can be attenuated by a further 15 dB, which is the factor granted 
by the square root of the number of spectral channels being averaged.

In summary, we find that a comprehensive approach which recognizes 
the inherent time, frequency and polarization structure of an RFI 
source allows for its exceedingly effective excision in the post-detection
stage when full Stokes data with suitable time and frequency 
resolutions are available.

\section*{Acknowledgments}
This work was supported by the Arecibo Observatory, 
which was then operated by Cornell University on behalf of the
National Science Foundation under a cooperative management agreement.

\end{document}